\documentclass[conference,10pt]{IEEEtran}

\usepackage{import}
\usepackage{packages}

\begin{document}
\title{Large-scale Outdoor Cell-free mMIMO Channel Measurement in an Urban Scenario at 3.5 GHz

\thanks{YZ and AFM are with the University of Southern California, Los Angeles, CA, USA; their work was financially supported by KDDI Research, Inc.. IK is with KDDI Research, Inc., Saitama, Japan. }
}

\author{
Yuning Zhang$^1$, 
Thomas Choi$^1$, 
Zihang Cheng$^1$, 
Issei Kanno$^2$, 
Masaaki Ito$^2$, \\
Jorge Gomez-Ponce$^{1,3}$, 
Hussein Hammoud$^1$, 
Bowei Wu$^1$, 
Ashwani Pradhan$^1$, 
Kelvin Arana$^1$, \\
Pramod Krishna$^1$, 
Tianyi Yang$^1$, 
Tyler Chen$^1$, 
Ishita Vasishtha$^1$, 
Haoyu Xie$^1$, 
Linyu Sun$^1$, \\
Andreas F. Molisch$^1$, \textit{Fellow, IEEE}\\

{\small $^1$ University of Southern California, Los Angeles, CA, USA, ~~~} \\
{\small $^2$ KDDI Research, Inc., Saitama, Japan, ~~}\\
{\small $^3$ ESPOL Polytechnic University, Guayaquil, Ecuador, ~~}
}

\maketitle

\begin{abstract}
    The design of \ac{CF-mMIMO} systems requires accurate, measurement-based channel models. This paper provides the first results from the by far most extensive outdoor measurement campaign for \ac{CF-mMIMO} channels in an urban environment. We measured impulse responses between over $20,000$ potential \ac{AP} locations and $80$ \acp{UE} at $3.5$ GHz with $350$ MHz \ac{BW}. Measurements use a ``virtual array'' approach at the \ac{AP} and a hybrid switched/virtual approach at the \ac{UE}.  This paper describes the sounder design, measurement environment, data processing, and sample results, particularly the evolution of the \acp{PDP} as a function of the \ac{AP} locations, and its relation to the propagation environment. 
\end{abstract}


\section{Introduction} 
        With the evolution of wireless communication systems from \ac{4G} to \ac{5G} and beyond, higher data rates and more uniformly reliable services are required\cite{tataria20216g}. \ac{CF-mMIMO} systems help to fulfill these requirements by distributing many \acp{AP} across a wide area, and connecting them all to a central processing unit using fast fronthaul connections; this reduces the average distance of a \ac{UE} from its nearest \ac{AP} and also reduces or eliminates the adjacent-cell interference \cite{zhang2019cell, demir2021foundations}, \cite[Chapter 22]{molisch2023wireless}. For an optimized system design, we need measurement-based channel models. Suitable measurements need to measure channels between many (tens to hundreds) distributed \acp{AP} and spread-out \acp{UE}. In particular, for system optimization, it is important to measure at all {\em potential} \ac{AP} locations, rather than at an inter-AP distance corresponding to the final deployment. 

        A few such measurements were conducted in the past in both indoor and outdoor environments. For the indoor case, measurements were done within limited indoor spaces (e.g., a single room), e.g., \cite{dey2019virtual, choi2020co, tawa2020measuring, li2022toward, perez2021experimental}. For outdoor environments, existing measurements, e.g., \cite{kurita2017outdoor, simon2023measurement, jungnickel2008capacity, hammons2008cooperative, maccartney2017base, loschenbrand2022towards}, were often intended for Cooperative Multipoint (CoMP) rather than CF-mMIMO and thus done with a small number of \acp{AP} ($\leq 32$, but often just $2-3$), which might be less than used in actual deployment of CF-mMIMO, and does not allow to extract statistics such as correlation of the shadowing at different potential \ac{AP} locations that are needed for versatile CF-mMIMO channel models. Our own previous work \cite{choi2022using} used a drone to form a large-scale virtual set of \ac{AP} locations. However, its relatively narrow \ac{BW}, low \ac{EIRP}, and poor phase stability due to the load limit of the drone and low movement stability, made it suitable only for extracting path loss (but not delay dispersion), and limited the dynamic range of the measurements.

        {\bf Contribution:} To overcome these limitations, this paper reports on the to-date most extensive outdoor CF-mMIMO channel measurement campaign. Specifically, we
        \begin{itemize}
            \item present an \emph{improved measurement technique} to implement a \emph{large-scale (210m-by-200m) virtual \ac{AP} array} combined with \emph{efficient solution} for \emph{multi-\ac{UE} measurement (switched operation)}; it can measure pathloss up to $140$ dB and achieves good phase stability;
            \item measured the channel between over \emph{20,000} potential \ac{AP} locations and \emph{80} \acp{UE} with \emph{350 MHz} \ac{BW} in an urban environment. The paper presents sample results of the \emph{\ac{AP-LD} channel impulse response}, along with its interpretation in terms of the \emph{ environment geometry}.
        \end{itemize}
        
        Our measurement results enable a more realistic approach to analyzing the impact of \ac{AP} density and placement, compared with the traditional ``power law with a superposed shadowing'' model that is widely used in theoretical research.

        {\bf Paper Organization: }A brief description of the channel sounder introduction and the measurement campaigns is given in Sec. \ref{sec: sounder system and measurement}. The evaluation procedures of the impulse responses are elaborated in Sec. \ref{sec: evaluation procedures} to show the process of obtaining the sample results that are then shown in Sec. \ref{sec: sample pdps}. We conclude the paper in Sec. \ref{sec: conclusion}.

\section{Sounder System and Measurement} \label{sec: sounder system and measurement} 
    \subsection{Sounder Setup}  \label{sec: sounder setup} 
\begin{figure}[htp] 
            \centering
                \includegraphics[width=1\linewidth, angle=0]{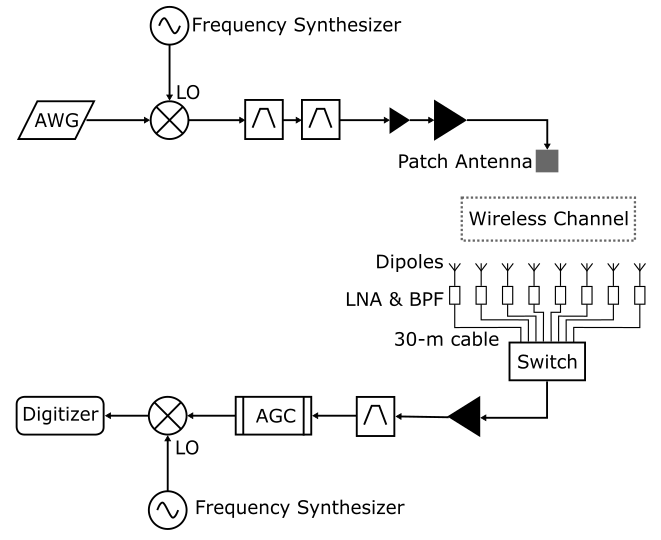}%
                \label{fig: block diagram}%
            \caption{Block diagram of the sounder systems. Minor parts, such as RF isolator and attenuators, are omitted. }
            \label{fig: system block diagram}
        \end{figure}

        The basic principle of the sounder combines a virtual array at the \ac{AP} with a hybrid switched/virtual array at the \ac{UE}. Specifically, the \ac{Tx} is placed on a cherry picker that is moved along a trajectory covering the potential \ac{AP} locations. A set of 8 \ac{UE} antennas is distributed over an area of $60$ m, and their signals are received by the \ac{Rx} via a fast electronic switch. The \ac{AP} repeats its route multiple times, with the set of \acp{UE} moved to different areas for each of those repetitions; the use of $N_{\rm UE}$ switched antennas thus reduces the number of repetitions of the trajectory by a factor of $N_{\rm UE}$. Fig. \ref{fig: system block diagram} shows the block diagram of the sounder; see \cite{zhang2024cell} for a more detailed sounder system description. 
        
        \subsubsection{\Ac{Tx}} 
            We use an \ac{AWG} to generate the \ac{IF} sounding waveform and have it transmit continuously. The generated waveform is up-converted to $3.5$ GHz \ac{RF} by a mixer with $3.875$ GHz from a frequency synthesizer (\ac{LO}), followed by two concatenated \acp{BPF} for \ac{LO} leakage - and upper-band - suppression. After being amplified by two concatenated \acp{PA}, a $42$-dBm signal is fed into a patch antenna\footnote{To be precise, into a switched dual-polarized patch antenna array, but only one element of the array, at vertical polarization, was used; for details see \cite{zhang2024cell}.}. The antenna is tilted $40^\circ$ downward, approximately facing toward the \acp{UE} on the ground level during the measurement, emulating the downtilt of typical \ac{AP} deployments. 
            
            The sounding waveform is a multi-tone signal that is similar to a Zadoff-Chu sequence used in LTE and 5G \ac{NR}. Our sounding sequence can maintain a flat spectrum and low \ac{PAPR} even after filtering and oversampling \cite{wang2017a}. A complete waveform takes $8\ \mu$s to transmit. With $2801$ sub-carriers, a $125$ kHz sub-carrier spacing is achieved, which in turn provides a maximum non-aliasing excess delay of $8\ \mu$s ($2400$ m); with $350$ MHz \ac{BW}, a $2.86$ ns ($0.86$ m) delay resolution can be achieved. 

        \subsubsection{\Ac{Rx}} 
            At the \ac{Rx}, we used $8$ dipole antennas as different \acp{UE}. Each of them is installed on a $1$-m high tripod and connected to the sounder \ac{Rx} via a $30$-m cable. The long cables enable uniformly distributing the \acp{UE} over an up-to-$60$m area. An $8$-to-$1$ switch cycles through collecting the signal from all \acp{UE} and passes them to the \ac{RF} chain. A customized \ac{AGC} unit follows, which is monitored and controlled by a National Instrument (NI) controller. The \ac{AGC} keeps the variation of the output power of the strongest (among the \acp{UE}) signal as low as $1.2$ dB\footnote{$1.2$ dB standard deviation over LOS or strong NLOS area, and $4.5$ dB over all locations. }, though the power from the other \acp{UE} may have bigger variations. The total \ac{NF} can increase by up to $20$ dB due to higher attenuator values in the \ac{AGC}, but as this occurs in regions of high received power, it does not impact the maximum (over the route) dynamic range of the setup ($\sim100$ dB). The output of the \ac{AGC} is finally down-converted to the \ac{IF} by the same \ac{LO} as the \ac{Tx} and sampled by a digitizer at $1.25$ GS/s.  

        \subsubsection{Switched array and sampling} 
            Both timing and frequency synchronization are precisely controlled by two GPS-disciplined \ac{Rb} clocks, which generate a $10$ MHz sinusoidal signal as a reference by all equipment, and a 1 \ac{PPS} as the start trigger for both sounders. 
            
            For each \ac{SISO} pair\footnote{We define one SISO pair as the radio link between the \ac{Tx} and one \ac{UE}. } the sounding waveform is repeated 10 times; the time interval between switching from one \ac{UE} to the next is $640 \mu$s, resulting in a measurement duration for one complete capture (one \ac{AP} location to 8 \acp{UE}) of $5.12$ ms. The \ac{Rx} turns to idle for transferring data to the storage device after such a capture and waits for the trigger for the measurement at the next \ac{AP} location, which occurs after $100$ ms. Given the speed of the cherry picker of $0.5$ m/s (see below), the interval between two measured \ac{AP} locations is approximately $5$ cm, while the movement of the \ac{AP} {\em} during a capture is only $0.2$ cm, i.e., a small fraction of a wavelength, and can thus be neglected.

    \subsection{Measurement Campaign} 
\begin{figure}
                \centering
                \includegraphics[width=.9\linewidth]{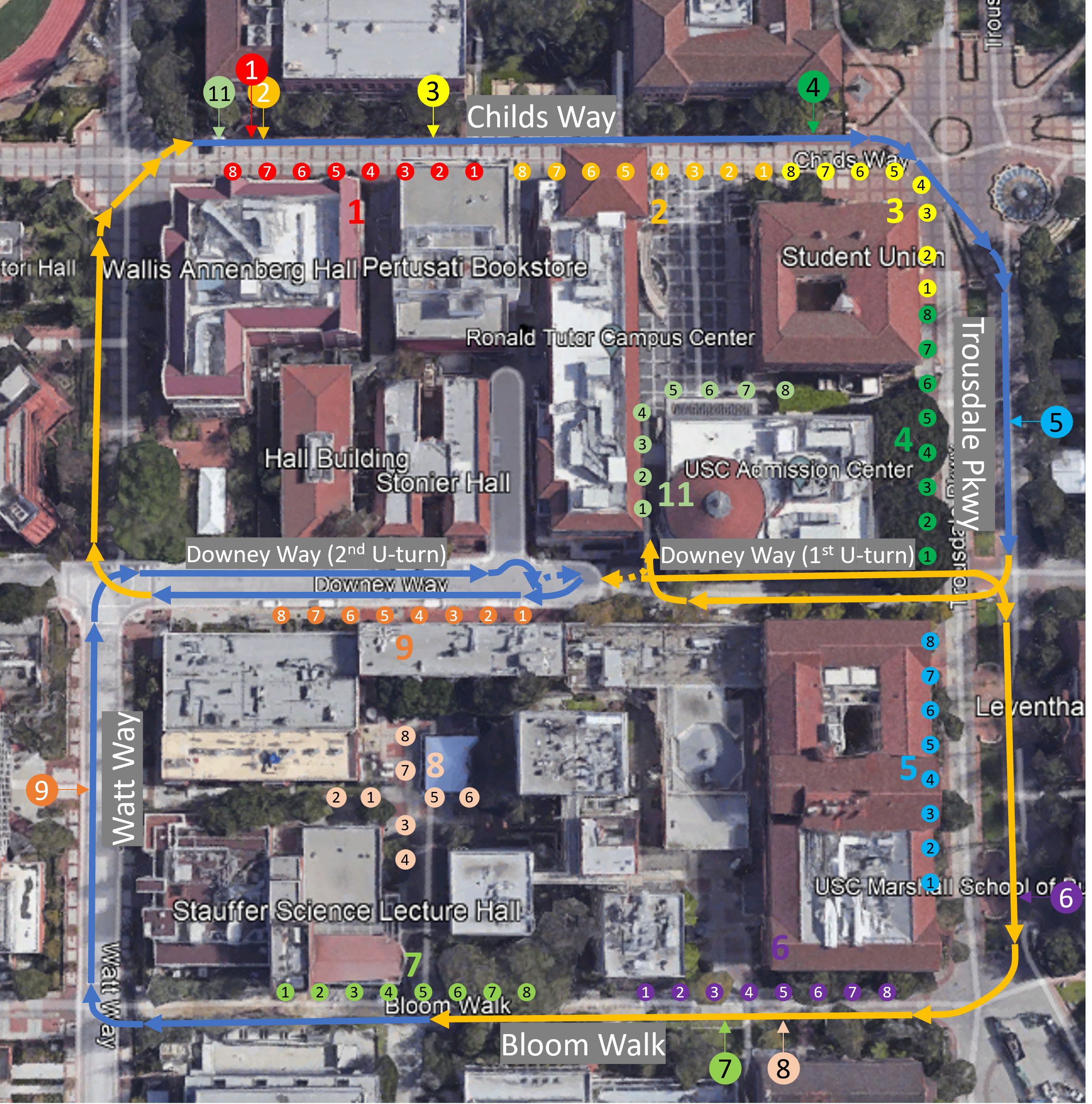}
                \caption{\small Small colored dots: measurement UE site locations, each color represents one UE cluster. Large colored circle with an arrow: start/stop location of the \ac{AP} trajectory for a particular \ac{UE} site. The ``north'' of the map is oriented $28^\circ$ clockwise from true North. Blue sections of the trajectory: AP height is $13$ m; yellow: $4\sim 5$m. Dashed parts: cherry picker geared in reverse.}
                \label{fig: measurement locations}
            \end{figure}

        The measurement campaign was conducted in an urban (but not metropolitan) environment, on the University Park campus of the \ac{USC}, Los Angeles, California, USA. We selected a $210 \times 200$ m rectangular area, with typical building heights on the order of $20$m; while the campus has a few high-rise buildings, none were contained in or near the measurement area. The \ac{AP} is installed on a precisely tracked heavy-duty cherry picker and moved at a speed of $0.5$ m/s following a certain trajectory to form the virtual \ac{AP} array. A total of $\NumUESiteTotal$ \ac{UE} sites\footnote{One \ac{UE} site covers an area that contains all $8$ spread \acp{UE}, which is centered with the \ac{Rx} sounder. Different \ac{UE} sites do not have overlapping areas. } are deployed (we re-form the large virtual \ac{AP} array for each \ac{UE} site). All $8$ \acp{UE} of the same UE site are spread out so that the $\NumUESiteTotal$ UE sites together cover much of the considered area; see Fig. \ref{fig: measurement locations} for the map of the measurement campaign location, \ac{AP} trajectory, and all UE sites. 

        Part of the \ac{AP} trajectory is $13$ m above ground, corresponding to typical heights of \acp{AP} \cite{3GPPTR38901}. However, in areas with trees, the height was lowered to $4\sim 5$m without changing the \ac{ULA} tilting angle\footnote{Due to the width of the elevation pattern, there is no need to adjust the tilting angle of the patch antenna while the \ac{AP} is low. The elevation pattern can be found in \cite{zhang2024cell}. }, both because \acp{AP} might be deployed at a lower height to avoid vegetation attenuation, and to protect the sounder from scratches and/or damage by the tree canopies. The cherry picker stops driving during the height change. The \ac{AP} is moved along the trajectory on the ``outside lane'' (with respect to the encircled street blocks) so that the \ac{ULA} is facing the deployed \acp{UE} across the streets by keeping the broadside of the \ac{ULA} pointing to the right side relative to the \ac{AP} movement direction. Note that the measurement location approaches (but does not perfectly emulate) actual \ac{AP} locations, which would typically be mounted flush on the building facades. The width of the streets vary between $3$ m and $15$ m. 

        To capture the environment (including possible time-varying scatterers) during the measurement, we use $360$-degree cameras to record the surroundings of the sounders throughout the entire measurement campaign. Caution tapes are also applied around the \acp{UE} to avoid pedestrians crossing and blocking the antennas.

\section{Evaluation Procedures} \label{sec: evaluation procedures} 

        We first apply \ac{FFT} to the captured time-domain data. Then, we performed averaging over $10$ repetitions of each \ac{SISO} capture, and we crop the band of interest and get it calibrated by the \ac{FFT} of the original sounding waveform to acquire the raw complex transfer function, $H_{{\rm raw},m,j,k}$, where $m$, $j$ and $k$ denote the index of the captures (time), \ac{Rx} antennas and frequency points. A \ac{B2B} calibration is performed as $H_{{\rm ch},m,j,k} = {H_{{\rm raw},m,j,k}} / {H_{{\rm cal},k}}$ to get the channel transfer function\footnote{Note that the calibrated transfer function includes the frequency response of the antenna patterns of both \ac{Tx} and \ac{Rx} antennas. While the antennas chosen for \ac{AP} and \ac{UE} are fairly typical, results do change when other types of antennas are used. Eliminating this impact would require double-directionally resolved measurements for each \ac{AP}/\ac{UE} pair, which is beyond the scope of this paper. }, where $H_{{\rm cal},k}$, the \ac{B2B} calibration data, is obtained by connecting the \ac{Tx} sounder output via a calibrated cable to the \ac{Rx} input individually (bypassing both \ac{Tx} and \ac{Rx} antennas), to each \ac{SISO} pair. We then filter it with a Kaiser-Bessel window ($\beta=3$) to provide delay-domain side lobe suppression though it does somewhat increase the main lobe width \cite{gomez2023impact}. A zero padding with a factor of $10$ is also applied to achieve oversampling in the delay domain.

        After taking the inverse \ac{FT} of the processed transfer function, we compute the \ac{PDP} from the \ac{CIR} as
            \begin{align}
                P_{{\rm h}_{m,j,\tau}} = \Big|\mathcal{F}^{-1}_{k} \big\{  H_{{\rm ch},m,j,k}  \big\}\Big|^2 ~,   \label{eq: PDP}
            \end{align}
        where $\tau$ is the index of delay bins regardless of oversampling or not, and $\mathcal{F}^{-1}_k\{\cdot\}$ denotes the inverse \ac{FT} operation over frequency points. 

        We subsequently apply noise thresholding to the $m'$th \ac{SSA} \ac{PDP}\footnote{We apply the small-scale averaging with a window size of $9$ snapshots ($0.5$ m of \ac{AP} movement), which is within the channel stationary region. } at a threshold level of $\theta_{m',j} = P_{{\rm n},m',j}+ \Delta_{\rm n}$, where $P_{{\rm n},m',j}$ denotes the average noise level and $\Delta_{\rm n}$ is a threshold of $7$ dB above this level \cite{gomez2023impact}. We also perform delay gating of the \ac{CIR}, retaining only the first $400$ delay bins because no significant \acp{MPC} can be observed beyond it. 
        
        The post-processing further eliminates observed cross-talk between the \ac{Tx} and the \ac{Rx} of the sounder, in that the transmitted signal couples directly in the \ac{Rx} circuitry, see detailed investigation process in \cite{zhang2024cell}. To remove such a cross-talk, we can set the pre-cursors to 0 in the post-processing from the first delay bin to the one that is $4$ bins ahead (the 4 bins are a safety margin) of the Euclidean distance regarding each individual \ac{UE} antenna.

\section{Measurement Results} \label{sec: sample pdps} 
    We now present sample results from our campaign, which show important trends for CF-mMIMO deployment and also demonstrate the propagation physics underlying them. More examples, as well as statistical evaluations of pathloss and delay spread, will be presented in \cite{zhang2024cell}. 
    
    Specifically, we show two different ``\ac{AP-LD} \acp{PDP}''. A complete \ac{AP-LD} \ac{PDP} is an ensemble of \acp{PDP} between all the \ac{AP} locations from a virtual \ac{AP} array and one particular \ac{UE}; they thus allow to assess the impact of changing the locations of the actual \acp{AP} and/or changing their density of deployment. Note that due to the virtual-array principle of our measurements, different measurement \ac{AP} locations (via the movement of the cherry picker) map to different times, i.e., the \emph{Time} axis in Fig. \ref{fig: AP-LD PDP 1} and \ref{fig: AP-LD PDP 2} is an indication of the AP locations. The sample results we show below are partial (to focus on the main effects) \ac{AP-LD} \acp{PDP} with $2$ different \acp{UE}. We will illustrate how clusters evolve as a function of continuously changing \ac{AP} location, which both provides insights into the dominant propagation processes and furthermore allows to verify whether propagation paths are plausible\footnote{We note that only delay information, but not angular information, is available. }. We also only focus on the \acp{MPC} that is within the $30$ dB dynamic range. 

    
    In Fig. \ref{fig: AP-LD PDP route 1} and \ref{fig: AP-LD PDP route 2} the colors of the trajectory, particularly green, orange, and blue, indicate the \ac{AP} locations where the channel is \ac{LOS}, \ac{OLOS}, and \ac{NLOS}, respectively (this color code is different from Fig. \ref{fig: measurement locations}, where colors denote AP height). The purple hexagram marks the \ac{UE} location. The beginning location\footnote{Strictly speaking, there is no ``beginning'' or ``ending'' location for actual \ac{AP} deployment. However, since different \ac{AP} locations in our measurement are implemented by the movement of the cherry picker, we then use such a concept to mark the two ends of the available \ac{AP} locations in the samples. } of the \ac{AP} in each sample \ac{AP-LD} \ac{PDP} is at the green triangle, and the ending point is at the red triangle. The \textcircled{R} and \textcircled{L} markers represent the locations where the \ac{AP} gets \emph{raised} and \emph{lowered} by the cherry picker. 
        
    \subsection{\Ac{AP-LD} \ac{PDP} Example 1} 
\begin{figure}[htp] 
            \centering
            \subfloat[AP-LD PDP 1]{%
                \includegraphics[width=1.1\linewidth, angle=0]{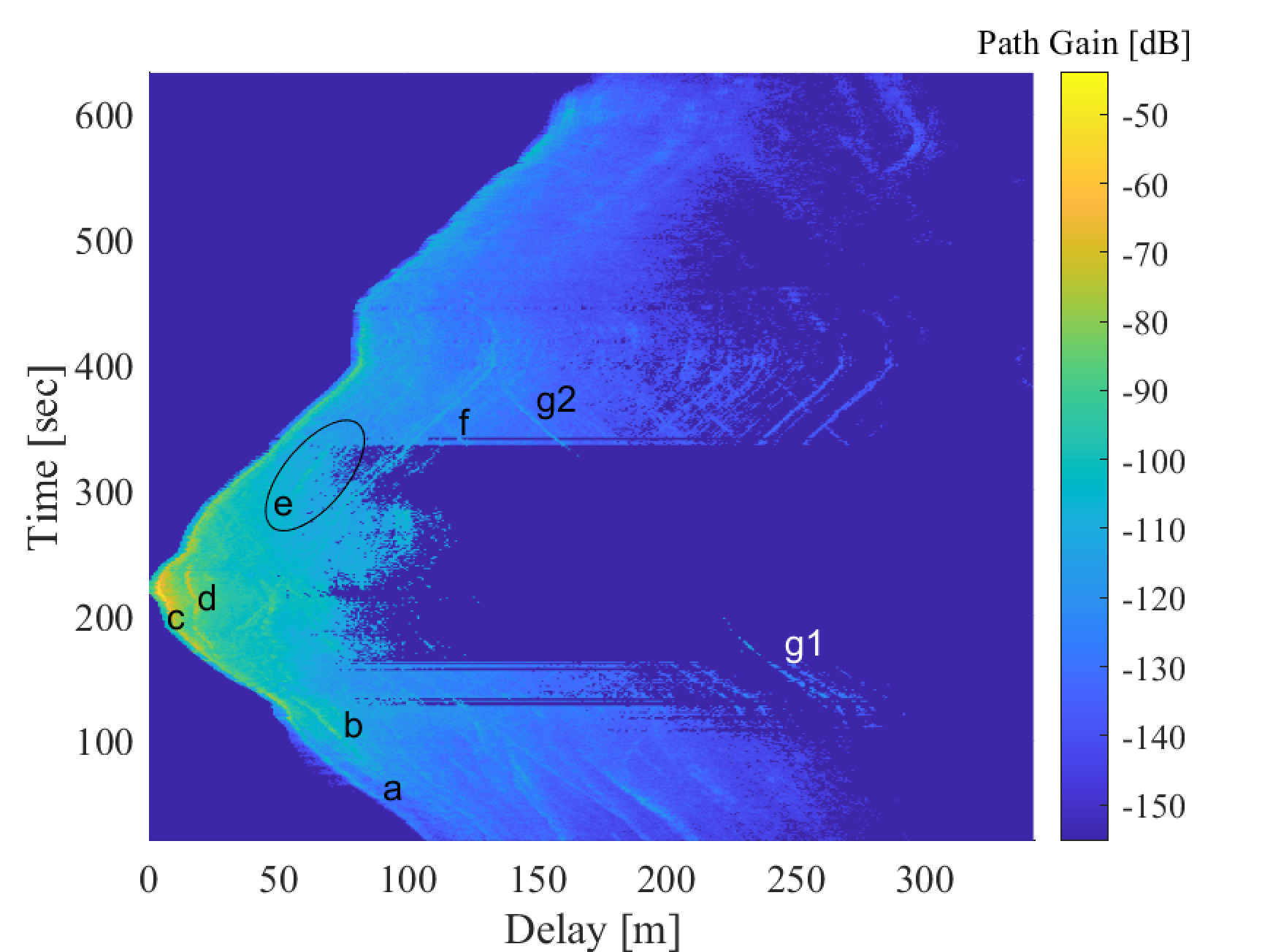}%
                \label{fig: AP-LD PDP 1}%
                }\\
            \subfloat[Measurement environment for AP-LD PDP 1]{%
                \includegraphics[width=1\linewidth, angle=0]{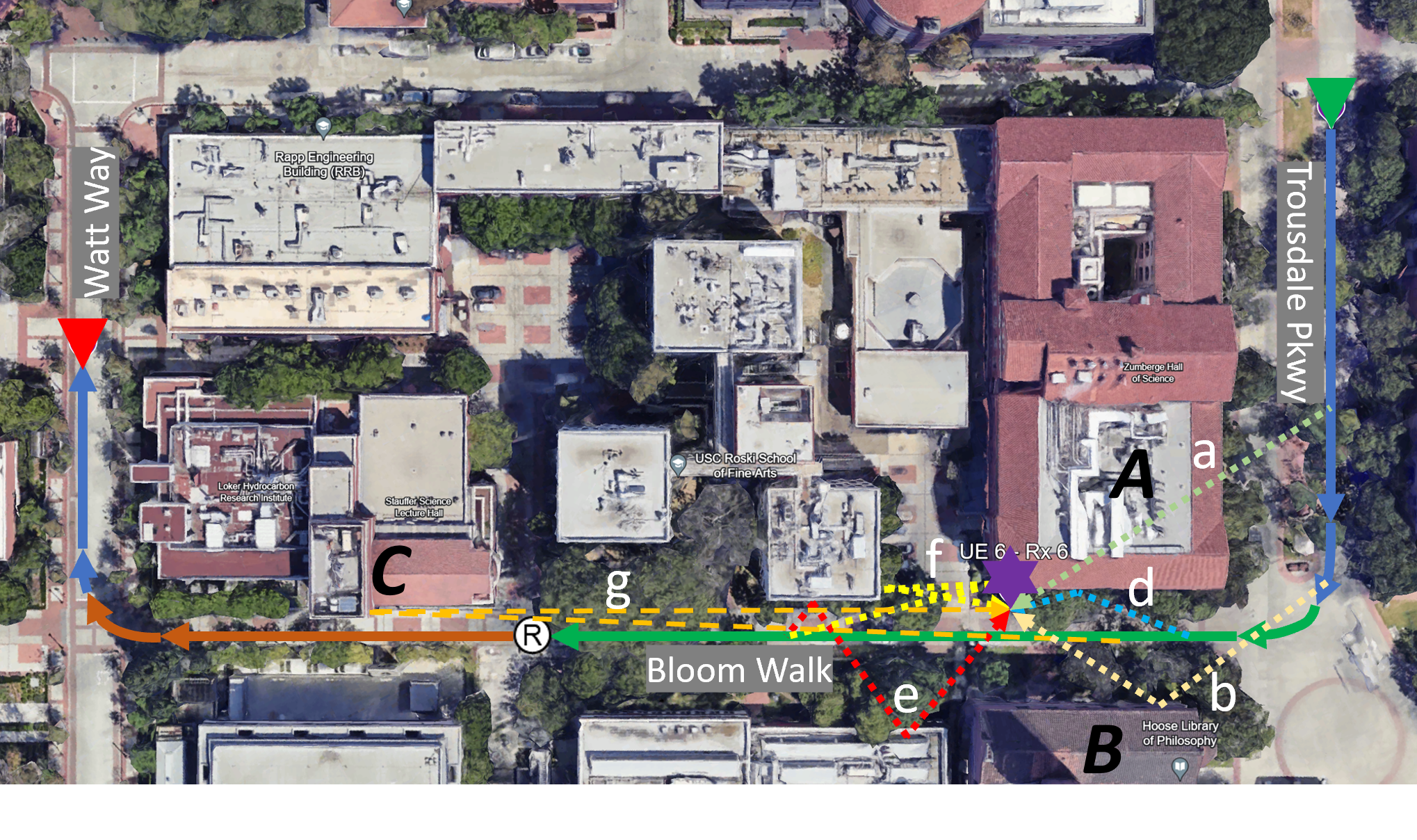}%
                \label{fig: AP-LD PDP route 1}%
                }%
            \caption{AP-LD PDP at UE site 6, Rx 6}
            \label{fig: AP-LD PDP example 1}
        \end{figure}

        The first \ac{AP-LD} \ac{PDP} example is shown in Fig. \ref{fig: AP-LD PDP 1}, and Fig. \ref{fig: AP-LD PDP route 1} shows the corresponding \ac{AP} trajectory overlaid with a Google Earth photo. The \ac{UE} is the $6$th antenna of the \ac{UE} site $6$, corresponding to a street canyon scenario. 

        The \ac{AP} starts out on Trousdale Parkway, in a \ac{NLOS} location: the excess delay of the first-arriving \ac{MPC} corresponds to propagation over the roof of the building $A$, which is noted by $a$. The Euclidean distance between the \ac{AP} and the \ac{UE} is shorter than $a$, but being invisible from the \ac{AP-LD} \ac{PDP} indicates it does not survive the noise thresholding, which in turn shows a high loss for the through-building propagation. Before the \ac{AP} turns into the Bloom Walk, a reflection occurs at building $B$, which generates the cluster $b$. Path $c$, between time $110$s and $400$s, is the direct \ac{LOS} link (corresponds to the green section in Fig. \ref{fig: AP-LD PDP route 1}, thus no particular propagation path is marked). Between time $160$s and $340$s, the absence of MPCs after about $100$ m is caused by a $20$ dB jump in the noise floor (and thus in the threshold for noise filtering) that is introduced by the \ac{AGC}. The cherry picker raises the \ac{AP} to the full height between time $400$s and $440$s. Since the height change is small compared to the distance between the \ac{AP} and the \ac{UE}, the delay of the earliest \ac{MPC} barely changes. 
        
        The channel becomes \ac{OLOS} while the \ac{AP} is raised due to the blockage of the tree canopies between the transceivers, and the dominant \ac{MPC} accordingly has reduced power. The cluster $d$ corresponds to a back-wall reflection while the \ac{AP} passes in front of the \ac{UE}. $e$ indicates the existence of a wave-guide effect in a street canyon scenario with multiple reflections from the buildings on both sides of the road. A two-bounce cluster, $f$, occurs after the \ac{AP} moves to the left side of the \ac{UE}, and lasts until being blocked by the trees during the raising process. The clusters $g1$ and $g2$ are conjectured to be caused by the long reflection from building $C$ because the delays at different times fit the physical distance. The ``missing'' part between $g1$ and $g2$ is likely caused by the above-mentioned change in threshold; in other words, $g1$ and $g2$ are caused by the same physical object and propagation process.



    \subsection{\Ac{AP-LD} \ac{PDP} Example 2} 
        \begin{figure}[htp] 
            \centering
            \subfloat[AP-LD PDP 2]{%
                \includegraphics[width=1.1\linewidth, angle=0]{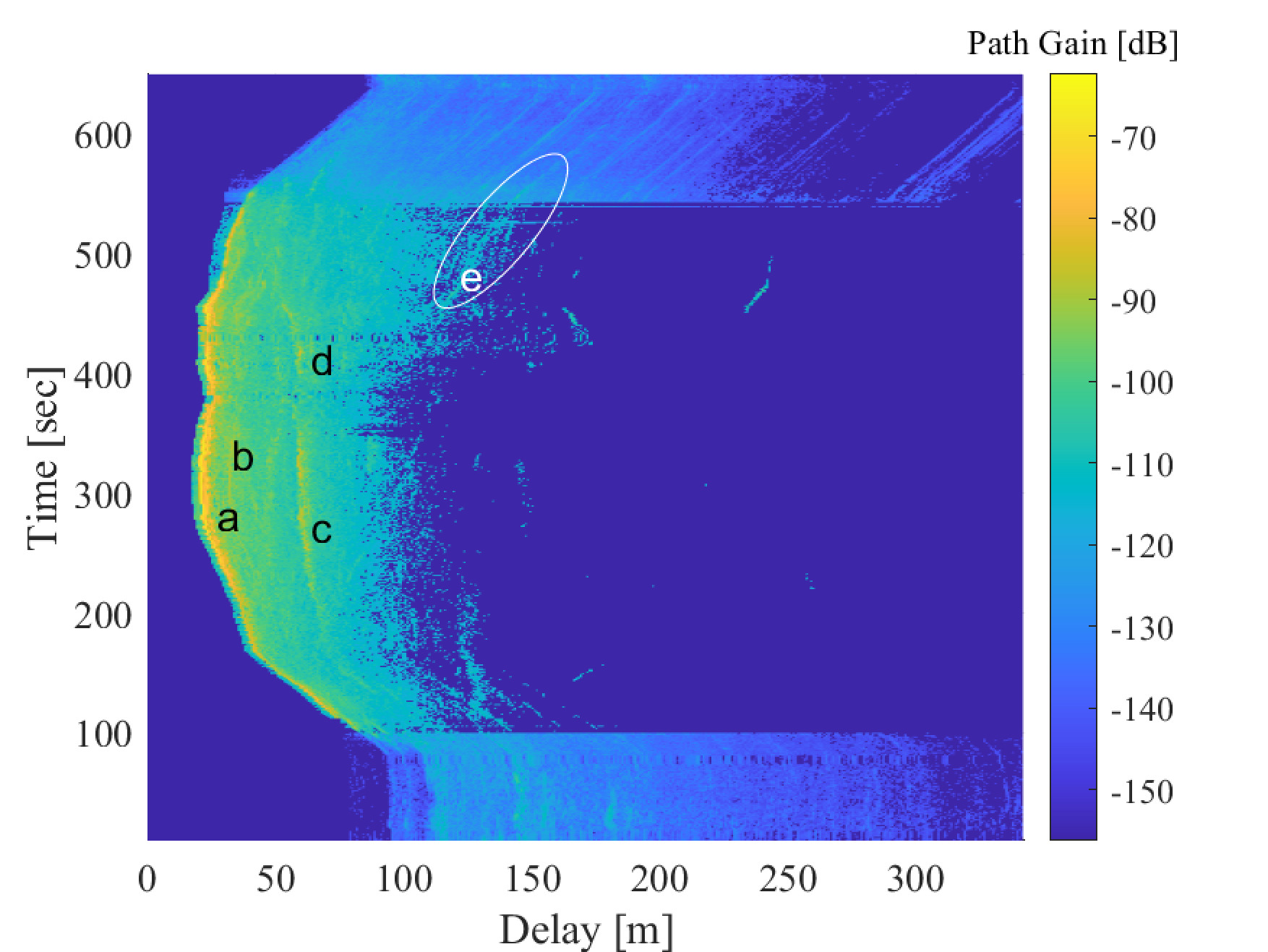}%
                \label{fig: AP-LD PDP 2}%
                }\\
            \subfloat[Measurement environment for AP-LD PDP 2]{%
                \includegraphics[width=1\linewidth, angle=0]{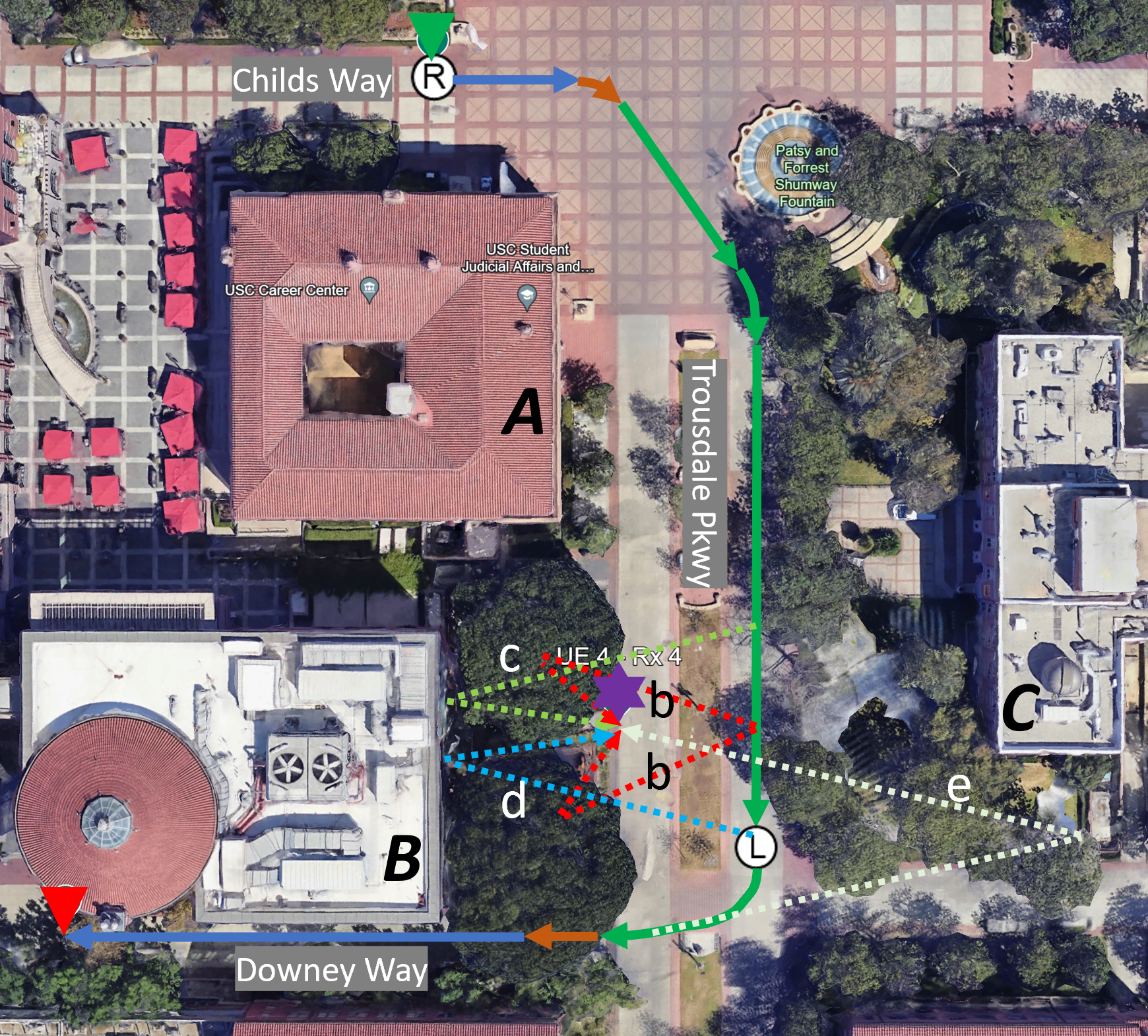}%
                \label{fig: AP-LD PDP route 2}%
                }%
            \caption{AP-LD PDP at UE site 4, Rx 4}
            \label{fig: AP-LD PDP example 2}
        \end{figure}
        The second example is taken from \ac{UE} site $4$, specifically the $4th$ \ac{UE} in it. The \ac{UE} is on a wider street and next to two large trees. Fig. \ref{fig: AP-LD PDP example 2} shows the \ac{AP-LD} \ac{PDP} and the corresponding map. The \ac{AP} is at the low position at the beginning of the \ac{AP-LD} \ac{PDP}. It gets raised to $13$ m before any lateral movement starts. 

        The first $90$s corresponds to the process of the cherry picker raising the \ac{AP}. After a short \ac{NLOS} and \ac{OLOS} section, a \ac{LOS} channel is available for a long time, from $110$s to $540$s, which is marked by the green section. During that time the noise figure and the threshold jump, similar to the effect described in AP-LD 1. Note that the AGC is governed by the strongest of the signals at the different \acp{UE}, and thus, the noise floor increases when {\em any} \ac{UE} in the chosen area has \ac{LOS}. Fig. \ref{fig: measurement locations} shows that all the \acp{UE} from the site $4$ (dark green) are on the Trousdale Parkway, and they share similar start and end moments of having \ac{LOS} links with the \ac{AP}. As soon as the \ac{UE} in this example has \ac{LOS} channel at time $110$s, the dominant \ac{UE} of the \ac{AGC} (the $8$th, which is at the top of the site) also has \ac{LOS} with strong received power, which makes the \ac{AGC} switch the ATT1 to the higher ($30$ dB) attenuation to keep the output power stable. When the \ac{LOS} channel ends at $540$s, the dominant \ac{UE} of the \ac{AGC} (the $1$st, which is at the bottom of the site) then loses \ac{LOS} as well, which makes the \ac{AGC} decrease the attenuation of ATT1, which decreases the noise floor correspondingly. 
        
        As the \ac{AP} changes locations and gets closer to the \ac{UE}, richer \acp{MPC} can be observed. Besides the \ac{LOS} path, the two most pronounced clusters are $b$ and $c$, which are caused by the reflection/scattering from the objects that are on the same side of the street as the \ac{UE} (as seen from the \ac{AP}). Particularly, cluster $b$ is caused by the foliage scattering when the \ac{AP} is in front of the \ac{UE}, and it has a short lifetime\footnote{We define the lifetime of a cluster as the duration that such a cluster exists in the \ac{AP-LD} \ac{PDP}.}, which indicates such scattering only takes effect when the \ac{AP} is at closer ranges. However, cluster $c$ has a much longer lifetime. It is caused by the reflection on the building $B$, after a small time when the \ac{AP} has \ac{LOS} connection with the \ac{UE}. The signal can penetrate the foliage from the sparse top and then get reflected. Between time $380$s and $470$s, the \ac{AP} is lowered, while remaining at location \textcircled{L}. During this time, the delay of the LOS component decreases again, since the AP now gets close to the UE due to {\em vertical} movement. At the same time, cluster $c$ disappears.  Instead, another cluster $d$ shows up after some gap time. The explanation for this gap is that the reflected path is temporarily blocked by the center of the foliage, which has the densest leaves and branches, causing high loss. After the \ac{AP} gets low enough, the reflected path can be reborn between the \ac{AP}, building $B$, and the \ac{UE}. Such effect provides insights on the height determination of the \acp{AP} when outdoor deployment is needed. 

        After the \ac{AP} changes its location to Downey Way, a reflected \ac{MPC} $e$ can be observed from the building $C$, and it vanishes while the \ac{AP} moves further, along with a lower gain on the side for the \ac{ULA} elements.


\section{Conclusions}   \label{sec: conclusion} 
    We have conducted the to-date most extensive outdoor \ac{CF-mMIMO} channel sounding measurement campaign with over $20,000$ \ac{AP} locations and $80$ \ac{UE} locations, using a combination of virtual array and switched-array technique. We presented the evolution of the clusters as a function of changing \acp{AP} locations. The sample results show the validity of our measurement data, by relating the dominant MPCs to the surrounding environment geometry. The sample results further show the various important propagation processes, including \ac{LOS}, over the rooftop propagation, building reflections, and vegetation attenuation. The results presented here, and the statistics from the campaign that will be presented in \cite{zhang2024cell}, provide important insights into the impact of environments and deployment density on the achievable performance of CF-mMIMO systems.

{\bf Acknowledgment:} This work was financially supported in part by KDDI Research, Inc., and in part by the National Science Foundation. JGP's work was financially supported in part by the Foreign Fulbright Ecuador SENESCYT Program.

\bibliographystyle{IEEEtran}
\bibliography{reference.bib}
\end{document}